# Evaluation of Four Black-box Adversarial Attacks and Some Query-efficient Improvement Analysis

Rui Wang
Department of Computer Science and Technology
Zhejiang University
Hangzhou, Zhejiang, China
rainwang6188@gmail.com

*Abstract*—With the fast development of machine learning technologies, deep learning models have been deployed in almost every aspect of everyday life. However, the privacy and security of these models are threatened by adversarial attacks. Among which black-box attack is closer to reality, where limited knowledge can be acquired from the model. In this paper, we provided basic background knowledge about adversarial attack and analyzed four black-box attack algorithms: Bandits, NES, Square Attack and ZOsignSGD comprehensively. We also explored the newly proposed Square Attack method with respect to square size, hoping to improve its query efficiency.

## I. INTRODUCTION

With the rapid development of artificial intelligence (AI), theories and technologies of deep learning have been widely applied to extensive range of areas, such as natural language processing, computer vision and decision making, etc. However, due to the poor interpretability of deep learning itself, the strong dependence on training data as well as the weak robustness of those models and algorithms, deep learning technology not only brings convenience to our life, but it also faces severe security problems, and adversarial attack is a typical attack method. In adversarial attacks, the fabricated samples, constructed by adding subtle perturbation to the original input, lead to model misclassifies or a wrong prediction. This paper will focus on adversarial attack in computer vision.

In this paper, we compare four famous black-box attack algorithms: Bandit [1], NES [2], Square attack [3] and ZO-signSGD [4], and focus on whether we can improve the query efficiency of square attack with respect to square size.
Before diving into the details of attack algorithms, some necessary background knowledges will be provided.

### A. Popular Image Datasets

Datasets are an integral part of the field of machine learning. Major advances in this field can result from advances in learning algorithms (such as deep learning), computer hardware, and, less-intuitively, the availability of high-quality training datasets.[5] Image datasets consisting primarily of images or videos for tasks such as object detection, facial recognition, and multi-label classification. Here are three popular image datasets used in object detection and recognition.

- MNIST: The MNIST database [6] of grayscale handwritten digits has a training set of 60,000 examples, and a test set of 10,000 examples. It is a subset of a larger set available from NIST. The digits have been size-normalized and centered in a fixed-size image.

- CIFAT10: The CIFAR-10 dataset [7] consists of 60000 32x32 color images in 10 classes, with 6000 images per class. There are 50000 training images and 10000 test images

- ImageNet: ImageNet [8] is an image dataset organized according to the WordNet hierarchy. Each meaningful concept in WordNet, possibly described by multiple words or word phrases, is called a "synset". There are more than 100,000 synsets in WordNet; the majority of them are nouns (80,000+). ImageNet provides on average 1000 images to illustrate each synset. Images of each concept are quality-controlled and human-annotated. In its completion, ImageNet will offer tens of millions of cleanly labeled and sorted images for most of the concepts in the WordNet hierarchy

### B. Convolutional Neural Network Models

Convolutional neural network (CNN) is a class of artificial neural network in deep learning, and is most commonly applied to analyze visual imagery like image recognition and classification [9]. The basic structure of CNN is represented in Figure 1, it consists of the input layer, convolutional layer, pooling layer, and a process of flattening, where the information is entered into a set of dense layers, representing the result obtained in the output layer.

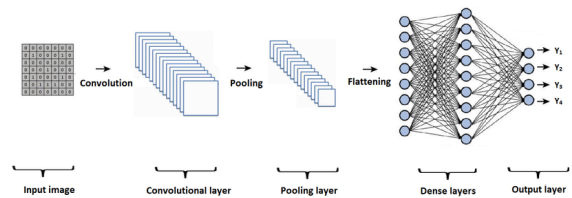

Fig. 1. Basic structure of a CNN.

*1) VGG16*

VGG16 is brought up by Simonyan and Zisserman [10] from Oxford. It is a simple convolutional neural network consists of 13 convolutional layers, 3 fully connected layers and 5 pooling layers, as is represented in Figure 2. The convolutional layers of VGG16 uses multiple 3×3 kernel-sized filter, which is better than those with one larger size kernel because multiple non-liner layers increases the depth of the network which enables it to learn more complex features at a lower cost.

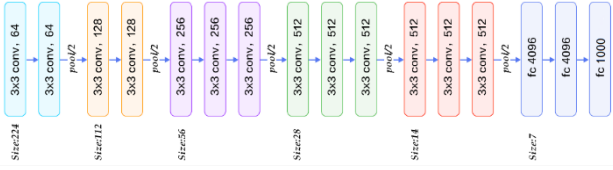

Fig. 2. Configuration of VGG16.

*2) Inception-v3*

Inception-v3 is a convolutional neural network architecture from the Inception family that makes several improvements including using Label Smoothing, Factorized 7×7 convolutions, and the use of an auxiliary classifier to propagate label information lower down the network. It has been proven that more than 78.1% accuracy can be achieved using ImageNet data sets [11]. The model is the product of many ideas proposed by several researchers over the years.

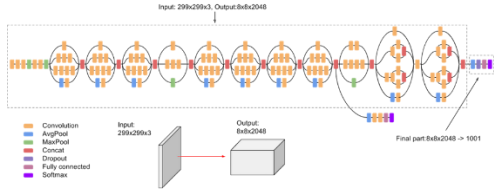

Fig. 3. Configuration of Inception-v3.

*3) ResNet-50*

ResNet stands for Residual Network. It is an innovative neural network that was first introduced by Kaiming He et al. in 2015[12]. Previous models use deep neural networks in which they stack many convolutional layers one after another since it was learnt that deeper networks perform better. However, it turned out that this is not always true. Deep networks have problems such as difficult to optimize, vanishing/exploding gradients and degradation problem. ResNet came up with the idea of skipping connections with the hypothesis that the deeper layers should be able to learn something as equal as shallower layers. They use the residual block to copy the activations from shallower layers and setting additional layers to identity mapping.

*C. Adversarial Attack*

Adversarial attacks pose a sever threat to the deep learning in practice, for it can completely deceive the neural network models simply by adding imperceptible perturbations to images.

According to the contexts, the threat model can be divided into white-box and black-box attacks. In white-box attack, adversary has all the privileges and complete access to the algorithm and execution, while in black-box attack, attacker has limited access to the parameters of the model and can only interact through input and output.

*1) White-box attack*

White-box attack implies that attackers have the ability to see into the inner workings of the network model, such as the architecture and the parameters. Most of current research findings employ white-box adversarial attack as the main method to generate adversarial samples. After analyzing most of the researchers' achievements, it is found that the white-box attack algorithms can be roughly categorized into the following four types [13]:

1. based on direct optimization
2. based on gradient optimization
3. based on decision boundary analysis
4. based on generative neural network (GNN) methods

*2) Black-box attack*

Contrast to white-box attacks, black-box attacker can only acquire restricted knowledge about the model, and which is closer to real-world scenarios. As a result, black-box attack usually execute by observing the outputs of queries like labels or confidence scores. According to Bhambri et al [14], black-box attacks can be divided into 4 categories based on the method they employ: gradient estimation, transferability, local search and combinatorics.

## II. ALGORITHM DETAILS

In this section, four algorithms will be covered in detail.

*A. Bandits*

Bandit optimization solves the multi-armed bandit problem in probability, a problem where a fixed limited set of resources must be allocated between competing choices in a way that maximizes the expected gain and only partial knowledge is known at the time of allocation. Andrew et al. employed the least square method in black-box adversarial attack to get the gradient estimator by solving a linear regression problem [1], and the author proved that this gradient estimator is approximately equivalent to those given by natural evolution strategy [2]. Moreover, they use the latent vector $v_t$ to incorporate the time-dependent priors and data-dependent priors in order to improve the query-efficiency. They cast the gradient estimation problem into the following bandit framework as described in Fig. 4, where $\mathcal{A}$ is the update rule, if the latent vector $v_t \in \mathbb{R}^n$, then $\mathcal{A}$ can simply use gradient ascent, where $\eta$ is the learning rate parameter.

```
Algorithm 1 Gradient Estimation with Bandit Optimization
1: procedure BANDIT-OPT-LOSS-GRAD-EST(x, y_init)
2:     v_0 ← 𝒜(φ)
3:     for each round t = 1, ..., T do
4:         // The loss in round t is ℓ_t(g_t) = − < Δ_x L(x, y_init), g_t >
5:         g_t ← v_{t−1}
6:         Δ_t ← GRAD_EST(x, y_init, v_{t−1}) // Estimated Gradient of ℓ_t
7:         v_t ← 𝒜(v_{t−1}, Δ_t)
8:     g ← v_T
9:     return Π_{∂𝒦}[g]
```

Fig. 4. Bandits Algorithm.

*B. Natural Evolution Strategies (NES)*

Natural Evolution Strategies (NES) update the search distribution in the direction of higher expected fitness using the natural gradient, which was first introduced into machine learning field by Amari in 1998 [15]. The use of the natural gradient makes the update independent on the particular parameterization of the distributions, resolving the limitations of plain search gradients.

NES are a family of evolution strategies, and the general procedure is as follows: a batch of search points are generated in the parameterized search distribution and the fitness function is evaluated at each point. The algorithm captures the structure of the fitness function via the distribution parameters. Then a search gradient on the parameters towards higher expected fitness is estimated by the algorithm. After that, NES performs a gradient ascent step along the natural gradient. The whole process repeats until meet a stopping criterion. The canonical NES algorithm is in Fig. 5, where $F$ is the Fisher information matrix of the given parametric family of search distributions. There're various techniques help to improve the performance as well as robust ness, like fitness shaping, adaptive sampling [2].

```
Algorithm 2 Canonical Natural Evolution Strategies
Input: f, θ_init
1: repeat
2:   for k = 0, ..., λ do
3:     draw sample z_k ~ π(·|θ)
4:     evaluate the fitness f(z_k)
5:     calculate log-derivatives Δ_θ log π(z_k|θ)
6:   Δ_θ J ← (1/λ) Σ_{k=1}^λ Δ_θ log π(z_k|θ) · f(z_k)
7:   F ← (1/λ) Σ_{k=1}^λ Δ_θ log π(z_k|θ) · Δ_θ log π(z_k|θ)^T
8:   θ ← θ + η · F^{-1} Δ_θ J
9: until stopping criterion is met
```

Fig. 5. Canonical NES Algorithm.

### C. Square Attack

Square Attack is a score-based adversarial attack method brought up by M. Andriushchenko et al. in 2020 [3]. It doesn't depend on the local gradient information and exploits random search, but Square Attack always uses the maximum perturbation budget at each step. The "Square" comes from the fact that the change in the image at each step are just small square-shaped contiguous pixels.

Square Attack employ two sampling distributions specific to the $l_2$- and $l_\infty$-attack, which are both motivated by the image procession in neural networks with convolutional filters as well as the shape of the $l_p$-balls for different p.

The general procedure of Square Attack is described in Fig. 6. It picks $h^{(i)}$ as the side length of the square, which is decreasing according to a predefined schedule. Then a new update δ is sampled in the distributions and added to the current iterate, after which the best loss is updated accordingly.

```
Algorithm 3 The Square Attack via random search
Input: classifier f, point x ∈ ℝ^d, image size w, number of color channels c,
       l_p-radius ε, label y ∈ {1, ..., K}, number of iterations K
Output: approximate minimizer x̂ ∈ ℝ^d that satisfies the optimization problem:
        min L(f(x̂, y)), s.t. ||x̂ − x||_p ≤ ε
1: x ← init(x), l* ← L(f(x), y), i ← 1
2: while i < N and x̂ is not adversarial do
3:   h^(i) ← side length of the square to modify
4:   δ ~ P(ε, h^(i), w, c, x̂, x) (the sampling distributions)
5:   x̂_new ← Project x + δ onto {z ∈ ℝ^d : ||z − x||_p ≤ ε} ∩ [0, 1]^d
6:   l_new ← L(f(x̂_new), y)
7:   if l_new ≤ l* then x̂ ← x̂_new, l* ← l_new
8:   i ← i + 1
```

Fig. 6. Square Attack Algorithm.

### D. ZO-signSGD

Zeroth-order sign-based stochastic gradient descent (ZO-signSGD) algorithm is introduced by Sijia Liu et. Al [4] in 2019. Since it uses zeroth-order gradient estimation, it is suitable for solving problems whose gradient is difficult or even infeasible to obtain. The generic sign-based SGD works as Fig. 7.

```
Algorithm 4 Zeroth-Order Sign-based Stochastic Gradient Descent
Input: learning rate {δ_k}, initial value x_0, and number of iterations T
1: for k = 0, 1, ..., T − 1 do
2:   ĝ_k ← GradEstimate(x_k)               ▷ see equation (2)
3:   x_{k+1} = x_k − δ sign(ĝ_k)            ▷ sign-gradient update
4: end for
```

Fig. 7. ZO-signSGD Algorithm.

## III. EXPERIMENT

In the experiment, the above four black-box attack algorithms will be compared according to their query efficiency and failure rate. Here only untargeted attacks is considered since the primary role is query efficient robustness evaluation. Note that we only cover the $l_\infty$-norm attacks.

First, we tested those black-box attacks on three DNN models with ImageNet dataset in terms of failure rate and query-efficiency for the $l_\infty$-attacks. Second, we explored whether the Square Attack is optimal with respect to square size $h$.

### A. Results for ImageNet

In this section, we compare the 4 black-box attack algorithms on the different model with different dataset. Fist, we thought about testing those attack method with CIFAR10 dataset with naturally trained models, but it appears all of the algorithms reach the nearly 100% success rate, maybe owing to the simplicity. As a result, we compare them on three pretrained models in Pytorch (Inception-v3, ResNet-50 and VGG-16-BN) using 1000 images from the ImageNet dataset. In the experiment, the suggested parameters are used as is described in their paper and we enforce a query limit of 10,000 queries for each point to find an adversarial perturbation of $l_\infty$-norm at most $\epsilon$=0.05.

The result is reported in Fig. 8. It's clear that the Square Attack has the lowest failure rate on all three models, and even gets 0.0% on ResNet-50 and VGG-16-BN. Moreover, the average number of queries of the Square Attack remains the lowest among the four black-box attacks. On the other hand, NES yields the highest failure rate and most average number of queries, largely owing to the computational overhead. For the rest of the two attacks, the difference is small, but Bandits may be relatively more query-efficient than ZO-signSGD.

| Norm | Attack | Failure Rate | | | Avg. Queries | | |
|---|---|---|---|---|---|---|---|
| | | I | R | V | I | R | V |
| $l_\infty$ | Bandits | 8.4% | 3.4% | 4.6% | 1339 | 854 | 596 |
| | NES | 13.2% | 8.7% | 6.5% | 1763 | 1335 | 918 |
| | Square Attack | **0.5%** | **0.0%** | **0.0%** | **217** | **78** | **31** |
| | ZO-signSGD | 10.6% | 7.8% | 2.2% | 927 | 887 | 687 |

Fig. 8. Summary of effectiveness of $l_\infty$-ImageNet untargeted attacks.

### B. Analysis of square size on query efficiency

As is stated in the previous section, Square Attack outperforms the other three black-box attack algorithms in failure rate as well as average queries. We're then interested in whether it is an optimal solution with respect to the size of the square. In the original scheme, the size $h$ of the square is calculated by the closest positive integer to $\sqrt{p \cdot \omega^2}$, where $p$ is the fraction of pixels to be modified, and $\omega$ is the length of the image, note that $p$ is the only free parameter of the scheme. At every iteration, only one square is generated in the image randomly with length h to employ the square attack, and the value of $p$ is halved at certain fixed iterations. As a result, we analyze the square size mainly on two aspects: generating multiple squares during each iteration and the modification of schedule of $p$. Several experiments are performed to verify the two factors.

*1) Results for multiple squares*

Our idea comes from the batch size in deep learning. Batch size is the number of samples that will be propagated through the neural network in one iteration. It is widely known that batch not only allows computational speedup from the parallelism of GPUs, but it also helps the model to achieve a better training stability as well as generalization performance [16]. So we wonder whether increase the number of squares during the square attack would benefit the query efficiency.

In the experiment, we randomly generate multiple squares and record the failure rate as well as the average queries. The result is shown in Table 1.

| Square Num. | Failure Rate | | | Avg. Queries | | |
|---|---|---|---|---|---|---|
| | I | R | V | I | R | V |
| 1 | 0.4% | 0.0% | 0.0% | 217.52 | 78.05 | 31.88 |
| 2 | 0.3% | 0.0% | 0.0% | 224.81 | 84.2 | 32.08 |
| 4 | 0.5% | 0.0% | 0.0% | 253.11 | 101.53 | 36.32 |
| 8 | 0.6% | 0.0% | 0.0% | 344.90 | 138.84 | 52.77 |
| 16 | 0.9% | 0.1% | 0.0% | 430.69 | 193.88 | 77.81 |

Fig. 9. Results of different square numbers in Square Attack on ImageNet.

We found that for ResNet-50 and VGG-16-BN, the original square attack scheme which attacks only one square already achieves the optimum failure rate, and increasing the number of squares to be attacked at each iteration simply increases the unnecessary number of queries. For the Inception model, although the original square attack scheme does not reach the optimum failure rate, increasing the number of squares benefits little to the failure rate (even increase the failure rate when square number is too large) but increase the average number of queries.

In general, attacking multiple squares on each iteration definitely increases the number of queries, but it has little positive influence to the failure rate.

*2) Results for different $p$ schedule*

In the original scheme, the percentage $p$ of elements of the image to be modified is based on a piece-wise constant schedule. Suppose the maximum iteration N=10000, then the value of $p$ will be halved at $i$ ={10, 50, 200, 1000, 2000, 4000, 6000, 8000} iterations, with N=10000 iterations available (with different N, the schedule is rescaled accordingly). This fixed constant schedule seems strange at first glance, so it is natural to assume that it's not optimal.

In the experiment, we arbitrary generate index list at which iteration p will be halved, and test the query efficiency. Here L1={10, 50, 200, 500, 1000, 2000, 4000, 6000, 8000} is the original index list, L2={10, 20, 50, 100, 200, 500, 1000, 3000, 6000} and L3={5, 40, 150, 400, 800, 1600, 2500, 4500, 8000} are the generated list. Note that L1 is sparse in the back while L2 is sparse in the front, and L3 is a small modification from L1. As is shown in Fig. 10, L3 yields a slightly better query efficiency than L1. We have also tested several randomly generated index list, we found that the original piece-wise constant schedule won't be optimal, but simply changing the index at which iteration p will be halved only has limited effect on query efficiency and failure rate.

| Index List | Failure Rate | | | Avg. Queries | | |
|---|---|---|---|---|---|---|
| | I | R | V | I | R | V |
| L1 | 0.4% | 0.0% | 0.0% | 217.52 | 78.05 | 31.88 |
| L2 | 0.4% | 0.0% | 0.0% | 271.55 | 91.93 | 33.54 |
| L3 | 0.3% | 0.0% | 0.0% | 235.53 | 74.45 | 30.50 |

Fig. 10. Results of different index list in Square Attack on ImageNet.

## IV. CONCLUSION

We have reviewed the basic knowledge of black-box attacks and compared four different black-box attack algorithms (Bandits, NES, Square Attack and ZOsignSGD). Among them Square Attack greatly outperforms the other three attack methods with respect to failure rate as well as number of queries. Some research about the square size of the Square Attack is performed to find out whether it reaches an optimum. We found that increasing the number of squares during each iteration results in little benefits to failure rate, but it significantly increases the average number of queries. As for the strategy of changing the value of $p$ (the fraction of pixels to be modified), the original schedule is certainly not the optimal but the benefit is quite limited. Other similar schedules yields approximately the same results.

Since $p$ is the only free parameter in the algorithm, there're still some space for exploration in the future. For example, the initial value of $p$ may depend on the training dataset, and we may come up a more query-efficient schedule for modifying $p$.